\documentclass[%
prd,
twocolumn,
superscriptaddress,
nofootinbib,
 amsmath,amssymb,amsfonts,
 aps,
showpacs
]{revtex4-1}

\usepackage{natbib}
\usepackage{url}
\usepackage{lineno}

\usepackage{bm}
\usepackage{dcolumn}
\usepackage{graphicx}
\usepackage{graphics}
\usepackage[latin1]{inputenc}
\usepackage{latexsym}
\usepackage{rotating}
\usepackage{xspace} 
\usepackage[usenames,dvipsnames]{color}
\usepackage{mathrsfs}
\usepackage{subfigure}
\usepackage{yfonts}

\usepackage{ulem}
\usepackage{outlines}
\usepackage{enumitem}
\setenumerate[1]{label=\Roman*.}
\setenumerate[2]{label=\Alph*.}
\setenumerate[3]{label=\roman*.}
\setenumerate[4]{label=\alph*.}

\definecolor {darkgreen}{rgb}{0.2,0.7,0.2}

\newcommand\be{\begin{equation}}
\newcommand\ba{\begin{eqnarray}}
\newcommand\ee{\end{equation}}
\newcommand\ea{\end{eqnarray}}

\newcommand\bw{\begin{widetext}}
\newcommand\ew{\end{widetext}}

\newcommand{\nn}{\nonumber}

\newcommand{\ppE}{{\mbox{\tiny ppE}}}

\newcommand{\ppN}{{\mbox{\tiny ppN}}}
\newcommand{\GR}{{\mbox{\tiny GR}}}

\newcommand{\WD}{{\mbox{\tiny WD}}}

\newcommand{\LISA}{{LISA}\xspace}
\newcommand{\SNR}{S/N\xspace}
\newcommand{\Mc}{\mathcal{M}}
\newcommand{\f}{f_0}
\newcommand{\fdot}{\dot{f}}
\newcommand{\fddot}{\ddot{f}}
\newcommand{\Msun}{M_\odot}

\begin{document}

\title{Binary White Dwarfs as Laboratories for Extreme Gravity with LISA}

\author{Tyson B. Littenberg}
\affiliation{NASA Marshall Space Flight Center, Huntsville, AL  35812, USA}

\author{Nicol\'{a}s Yunes}
\affiliation{eXtreme Gravity Institute, Department of Physics, Montana State University, Bozeman, MT 59717, USA.}

\date{\today}


\begin{abstract} 

The observation of low-frequency gravitational waves with the Laser Interferometer Space Antenna will allow the study of new sources of gravitational radiation that are not accessible by ground-based instruments. 
Gravitational wave sources provide invaluable information both about their astrophysics, as well as the nature of the gravitational interaction in their neighborhoods. 
One low frequency source that has not received much attention regarding the latter are galactic binaries composed of two white dwarves or a white dwarf and a neutron star. 
We here show that, contrary to the common lore, such gravitational wave sources can indeed be used to constrain an important feature of the gravitational interaction: the absence of pre-Newtonian, dipolar dissipation.
We propose a model-independent framework to carry out a null test for the presence of this feature in the data that is very much analogous to tests of General Relativity with radio-observations of binary pulsars.
We then go one step further and specialize this test to scalar-tensor theories to derive projected constraints on spontaneous scalarization. 
We find that these constraints can be comparable to current bounds with binary pulsars, and in some optimistic cases, they can be even stronger. 

\end{abstract}


\maketitle

\section{Introduction}

Sooner than all but the most hard-line optimists would have dared to imagine, gravitational waves (GWs) were observed by the advanced LIGO detectors in August 2015~\cite{Abbott:2016blz,Abramovici:1992ah, Abbott:2007kv, Harry:2010zz, Abramovici:1992ah}, forever changing physics and astronomy. 
That momentous discovery single-handedly confirmed the reality of GWs and their direct interaction with matter, an issue that although taken for granted by the relativity community, was still contentious in some astrophysical circles. And at that point, the flood gates opened. Discovery after discovery was heralded with great enthusiasm, leading to the first triple-coincident discovery when the advanced Virgo detector joined the network~\cite{Caron:1997hu, Giazotto:1988gw, TheVirgo:2014hva, virgo,Abbott:2017oio}, and later the first coincident discovery when light from across the electromagnetic spectrum, from gamma ray telescopes in orbit to radio telescopes on the ground, was detected in an unprecedented, coordinated observing campaign~\cite{TheLIGOScientific:2017qsa, Monitor:2017mdv, Tanvir:2017pws}. 

The era of GW astronomy has thus begun, and with it, great discoveries and some surprises have trickled in from the interpretation of the observations made. These discoveries can be grossly classified into two separate categories: astrophysics and fundamental physics. On the astrophysics side, we have learned that the most common sources of GWs are the coalescence of black holes in quasi-circular inspirals, and perhaps the biggest surprise is the black hole masses detected. Before the GW era, the heaviest stellar-mass black holes observed were less than ${\sim}16 \Msun$~\cite{Orosz:2007ng}, but several of the GW observations are consistent with relatively heavy, stellar-mass black holes merging into objects of above $40 \Msun$~\cite{Abbott:2016blz,Abbott:2016nmj}. Naturally, this surprise had important implications on the astrophysics of black hole formation, restricting certain models and supporting others~\cite{TheLIGOScientific:2016htt}. 

On the fundamental physics side, the GW era has done much more than simply confirm the existence of GWs: it has allowed us to explore, for the first time, the \emph{extreme gravity} regime, where the gravitational interaction is phenomenally large and dynamically changing, matter transcends into new forms not found on Earth, and physics is intrinsically non-linear and dynamical~\cite{Yunes:2013dva}. The latest GW observation of merging NSs has allowed for the first stringent constraints on the mass and radius of NSs, and thus on the equation of state of supranuclear matter~\cite{Annala:2017llu, Radice:2017lry,Abbott:2018exr,Chatziioannou:2018vzf,Yagi:2015pkc,Yagi:2013bca,Yagi:2013awa}. These constraints, in turn, have important implications on the theory of nuclear interactions at high densities, providing support for certain approximate methods to solve the many body Schr\"odinger equation~\cite{Chatziioannou:2015uea}. Moreover, this same observation, together with electromagnetic counterparts, has allowed for the first constraints on the speed of GWs, and thus, the speed of gravity~\cite{TheLIGOScientific:2017qsa,Monitor:2017mdv,Yunes:2016jcc}. These constraints, in turn, have dramatic implications on theories that modify General Relativity (GR) to explain late-time cosmological observations or to unify GR with quantum mechanics~\cite{Yunes:2016jcc}. 
 
These first advanced LIGO/Virgo discoveries, however, are only the tip of the iceberg. Current ground-based detectors are expected to be upgraded to third-generation designs, allowing for detections from effectively everywhere in the universe where stellar-mass or intermediate-mass black hole collisions occur. But even with this technology, such detectors will be limited to observing GWs with frequencies above a few Hz due to inescapable Newtonian noise. The observation of low frequency GWs will therefore require different measurement techniques. Access to the low-frequency band between ${\sim}0.1$ and ${\sim 10}$ mHz, where the richest population of sources is expected, requires a space-based GW observatory: enter the Laser Interferometer Space Antenna~\cite{Prince:2003aa,Audley:2017drz}, or \LISA for short.    

By opening up the low-frequency band, \LISA will allow for a plethora of new astrophysical and fundamental physics discoveries because it will give us access to entirely different classes of sources: supermassive black hole coalescences, extreme mass-ratio inspirals, and ultra compact binaries, typically comprised of white-dwarf/white-dwarf or neutron-star/white-dwarf pairs, in our galaxy. The observation of supermassive black holes will reveal invaluable information about structure formation in the Universe, including how galaxies collide and form, as well as how black hole remnants relax to a stationary configuration after a major collision with another black hole~\cite{Gair:2012nm}. GWs produced as a small compact object zooms and whirls into a supermassive black hole, a so-called extreme mass-ratio inspiral, will allow for a geodesic mapping of the spacetime created by the supermassive object, allowing for tests of the Kerr hypothesis~\cite{Gair:2012nm, vigelandnico,gairyunes}.   

But what about GWs emitted by galactic binaries? Binaries composed of two white dwarves (WDs) or a WD and a NS in the Milky Way are known to exist because they are observed electromagnetically. Several of these known sources are guaranteed to be detectable for \LISA, and are thus referred to as \emph{verification binaries}~\cite{Stroeer:2006rx,Korol:2017qcx, Kupfer:2018jee}. Moreover, given such observations and using population synthesis simulations, it is predicted that tens of millions of such binaries are emitting GWs in the \LISA band. Tens of thousands of these will be individually detectable by \LISA, while the rest blend together to form a confusion-limited foreground that is the dominant source of noise in the \LISA band between $0.4$ mHz and $2$ mHz~\cite{Benacquista:2006sc,Cornish:2017vip}. 
The catalog of resolvable binaries, as well as the spectral shape of the confusion noise, will contain a trove of new information about the formation and evolution of compact binaries in the galaxy, and the complicated physical interactions between binaries including mass transfer, common envelop interaction, and the internal physics of compact stellar remnants~\cite{Takahashi:2002ky,Crowder:2004ca,Barack:2004wc,Umstatter:2005jd,Umstatter:2005su,Cornish:2005hd,Rubbo:2006nq,Kopparapu:2006fi,Benacquista:2006sc,Benacquista:2007ct,Littenberg:2011zg,Cornish:2017vip}.

But could such sources also be used to learn something new about gravity itself?  
The expected answer to this question is a rotund no and the lore is cemented on (at least) the following three arguments. 

\emph{Theoretical Physics Argument}: GW-emitting galactic binaries detectable by \LISA will be well-separated, with distances of roughly $10^{5}$ km for typical WD masses at GW frequencies of $(10^{-2},10^{-3})$ Hz.  Such large separations imply orbital velocities smaller than $10^{-2} c$, which is certainly in the regime of velocities accurately probed by binary pulsar observations. Because binary pulsars have verified GR to amazing accuracy, modifications to gravity are expected to only be important (if they arise at all) at much higher velocities or smaller orbital separations. If so, then there is no point in using galactic binary signals to test GR since these will be probing a regime of gravity that has already been tested to a great degree by binary pulsar observations.

\emph{Data Analysis Argument}: Such small separations imply orbital periods of tens of minutes, which in turn imply an extremely slow chirping rate of $\dot{f} < 10^{-15} \; {\rm{Hz}}^{2}$. This is why galactic binaries are sometimes said to be \emph{monochromatic} sources of GWs, emitting mostly at a single frequency during the observation period. Thus, such GW signals contain less accessible information than the familiar merger signals from ground-based observatories, since the former cannot be used to measure higher frequency derivatives, which encode information about the mass ratio and the spin of the binary components. The limited information contained in galactic binary signals implies that they would lead to suboptimal tests of GR. 

\emph{Astrophysical Argument}: Consider the effect of astrophysical processes on the orbital evolution of galactic binaries. WDs are considerably more deformable than NSs due to their weaker gravitational compactness. As such, WDs in a binary can easily be deformed if the orbital separation is small enough. What is worse, if the separation is even smaller, a WD in a binary will transfer its mass to a more compact companion, once its Roche Lobe is filled, or there could be mass transfer through winds. All of these astrophysical effects can contaminate or even control the orbital motion, and thus, the GW emission. 

Putting these three arguments together (theoretical, data analysis and astrophysical) has led the field to conclude that galactic binaries would not be good laboratories for extreme gravity experiments, but is this really the case? Let us show through a simple two-fold argument that there is important information contained in the GWs emitted by galactic binaries that could be used to gain new knowledge about gravity.

Our first counter-argument is related to astrophysics. It is definitely true that WDs can be deformed, and their orbital evolution can be contaminated by mass transfer and/or tidal effects, as presented in the astrophysical argument above. However, higher mass WD binaries are more compact, and they can reach smaller orbital separations before interactions between the binary components become dominant in the evolution. High mass binaries are also among the loudest \LISA sources anywhere in the galaxy, making them excellent candidates for detailed parameter estimation studies. While a small subset of the total number of detectable binaries in the galaxy, the sheer volume of sources suggests that there will be a few exceptional systems in the high mass tail of the population that can be exploited as laboratories to test gravity.

Our second counter-argument is related to theoretical physics. Again, it is definitely true that extreme gravity modifications to GR are expected to be dominant in the late inspiral, and thus, be suppressed in the low velocity regime, as presented in the theoretical physics argument above. However, there are some (very popular) modifications to GR that do exactly the opposite: they are dominant at low velocities. Such modifications are sometimes called ``pre-Newtonian,'' because they enter as negative powers of the velocity relative to the leading-order GR expression. A popular example is dipole radiation through the activation of a scalar or vector field, as in the case of scalar-tensor theories~\cite{Damour:1992we,Damour:1993hw}, quadratic gravities theories, like dynamical Chern-Simons gravity~\cite{CSreview,jackiw,Campbell:1990fu,Campbell:1990ai,yunespretorius,kent-CSBH,quadratic,Yagi:2013mbt,Stein:2013wza} or Einstein-dilaton-Gauss-Bonnet gravity~\cite{1992PhLB..285..199C,Kanti:1995vq,Alexeev:1996vs,Torii:1996yi,Yunes:2011we,Ayzenberg:2014aka,Stein:2013wza,Yagi:2015oca}, and Lorentz-violating theories, like Einstein-\AE{}ther theory~\cite{Jacobson:2000xp,Eling:2004dk,eling-AE-NS,Yagi:2013qpa,Yagi:2013ava} and Horava gravity~\cite{Horava:2009uw,Blas:2009qj,Yagi:2013qpa,Yagi:2013ava}.

The emission of dipole radiation through the activation of new degrees of freedom forces the binary to decay,  and therefore the GWs to chirp, faster than in GR. For example, the rate of change of the GW frequency in GR scales as $v^{-11} \mu/m^{3}$, where $v$ is the orbital velocity, and $\mu$ and $m$ are the reduced and total masses of the binary. But in theories with dipole radiation, the chirping rate acquires a correction that scales as $v^{-13} \mu/m^{3}$, which is thus dominant when $v \ll 1$ and invalidates the theoretical physics argument above. The observational implication for galactic binaries is thus clear: if pre-Newtonian dissipation is present, the GW chirping rate will be at least four orders of magnitude larger than the GR prediction, leading to non-monochromatic galactic binary signals in the \LISA band. The absence of such an effect should thus allow us to place stringent constraints on the existence of pre-Newtonian modifications to GR.

With the theoretical physics and the astrophysical arguments counter-acted, we are then left only with the data analysis argument, which we recall states that there is simply not enough information in a galactic binary GW signal to carry out a test of GR. Whether this is true or not requires a data analysis study, as it is hard to assess otherwise. We will here show through such a study that, in fact, the data analysis argument is unfounded and tests of GR are indeed possible. Such a study allows us to develop both a model-independent and a model-dependent framework to test GR. 

The model-independent framework is a null test of GR and, in fact, it is very much analogous to complementary tests that have been carried out with radio-observations of binary pulsars. The main idea is to extract the GW signal with a model parameterized only by its overall amplitude, frequency, first and second frequency derivatives, while remaining agnostic about the particular source that produced it. In particular, this means we impose no prior between the frequency derivatives and the mass of the system or frequency of the signal when we do parameter estimation. The best fit model parameters can then be used to draw curves in the mass-frequency plane assuming GR is correct, and the test is passed if all curves intersect at the same location.  

For such a model-independent test to work, one must first detect at least one galactic binary that is sufficiently bound so that its second derivative is measurable, while at the same time not so bound that its evolution is controlled by mass transfer. From a galaxy simulation consistent with those used for past \LISA performance studies and mock data challenges~\cite{Nelemans:2003xp}, we select the ten highest frequency detached (i.e.~non-interacting) binaries to study the \LISA measurement of orbital evolution, and how those results can be used to study gravity theories.  

\begin{figure}
	\includegraphics[width=0.5\textwidth]{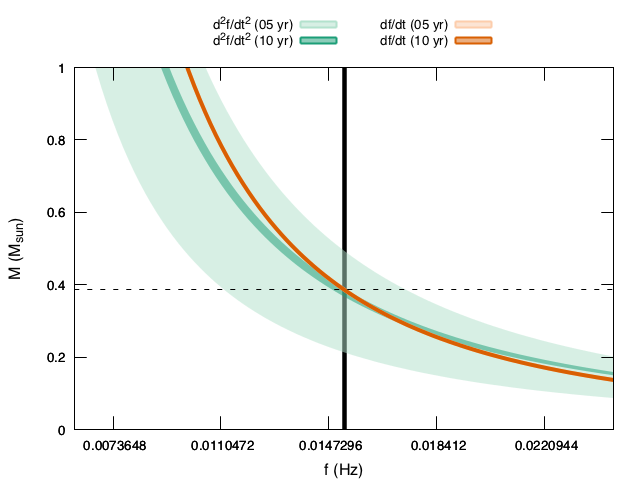}
	\caption{(Color online) Null test of GR with WD binaries. Each extracted parameter (the frequency (black), the first (orange) and the second frequency derivatives (green)) are curves in the chirp mass-frequency plane. If GR is correct, as in the example shown here for a 5 year (light shade) and 10 year (dark shade) observation, all curves intersect in the same region. Note that differences in the 5 and 10 year curves inferred from the first frequency derivative are not noticeable in this figure.}
	\label{fig:Mc_f.png}
\end{figure}

Figure~\ref{fig:Mc_f.png} presents an example of this null test in the chirp mass--frequency plane, where the former is a particular combination of the component masses of the system. The measured GW frequency is a vertical curve in this plane (black in the figure), whose thickness represents its $90\%$ credible interval. Assuming the null hypothesis that GR is correct, the inferred value of the first frequency derivative can be used to draw another curve in this plane (orange in the figure), where again its thickness is its $90\%$ credible region. The intersection of these two curves gives the inferred chirp mass of the system assuming the null hypothesis. Any additional information is thus redundant and can be used to test the null hypothesis. For example, the measured value of the second frequency derivative can be used to draw another curve in this plane (green in the figure), and if the null hypothesis is correct, then all curves intersect in the same region.   

In addition to GR consistency tests, galactic binaries can also be used to test specific models that predict the existence of pre-Newtonian dissipation. One such model is massless scalar-tensor theory, perhaps one of the simplest modifications to GR~\cite{Brans:1961sx,Damour:1992we,Damour:1993hw,Damour:1996ke,Damour:1998jk}. In this theory, a massless scalar field couples to the metric tensor, thus affecting the structure of massive objects and their motion in strong gravitational fields. In one sub-class of such models~\cite{Damour:1992we,Damour:1993hw,Damour:1996ke,Damour:1998jk}, Solar System observations can be easily passed because the scalar field is shielded and does not activate in weak gravitational fields. In strong gravitational fields, such as those produced by NSs and WDs, the scalar field does activate, producing pre-Newtonian dissipation effects~\cite{Will:1989sk,Damour:1998jk}. Therefore, constraints on pre-Newtonian dissipation could be used to place bounds on the two coupling constants of this class of theories $(\alpha_{0},\beta_{0})$. 

We here study the strength of such bounds using GWs emitted by galactic binaries of the mixed WD-NS type. The pre-Newtonian dissipation in scalar-tensor theories does not only depend on $(\alpha_{0},\beta_{0})$, but also on the response of the star to the presence of a scalar field, quantified in terms of certain sensitivities, which in turn depend on the compactness of the stars. Such tests, therefore, require first the independent measurement of the masses of the stars, which could be obtained electromagnetically if the binaries are eclipsing~\cite{parsons}. These masses can then be used to infer the radius of the stars, and thus their compactness, using knowledge of the stellar equation of state, which for neutron stars we assume will be known by the time \LISA flies through a combination of advanced LIGO/Virgo observations~\cite{Annala:2017llu, Radice:2017lry,Abbott:2018exr,Chatziioannou:2018vzf,Yagi:2015pkc,Yagi:2013bca,Yagi:2013awa} and X-ray observations~\cite{Miller:2014mca,Ozel:2015ykl}. 

\begin{figure}
	\includegraphics[width=0.5\textwidth]{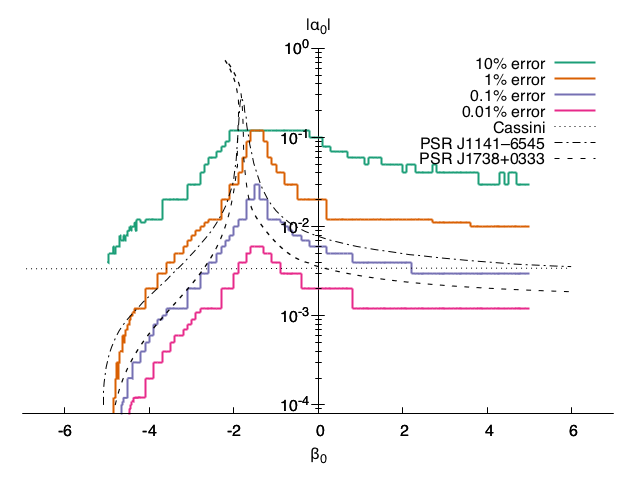}
	\caption{(Color online) Constraints on the $(\alpha_{0},\beta_{0})$ coupling parameters of scalar-tensor theories with \LISA observations of GWs emitted by WD-NS-star binaries, assuming different accuracies in the electromagnetic determination of the NS mass. For comparison, we also show constraints from radio-observations with a couple of binary pulsars~\cite{kramer-wex,Verbiest:2012ie}, as well as Solar System constraints from the tracking of the Cassini spacecraft~\cite{Bertotti:2003rm}. Observe that \LISA constraints are comparable to those set by binary pulsars when the component masses can be measured with high precision.}
	\label{fig:alpha_beta.png}
\end{figure}

With these assumptions, a GW measurement of $(\f,\dot{f},\ddot{f})$ implies a constraint of an open region in the $(\alpha_{0},\beta_{0})$ plane. Such constraints would be very much analogous to those placed with radio-observations of binary pulsars~\cite{Stairs:2003eg,Freire:2012mg}. We indeed find that the constraints \LISA could place in the future will be complementary, comparable and sometimes better than what binary pulsars can do today. Figure~\ref{fig:alpha_beta.png} shows the constraints that \LISA could place assuming different accuracies for the measurement of the masses and radii of the component stars. Observe that the $(\f,\dot{f},\ddot{f})$ constraints are comparable to those that can be placed with radio-observations of binary pulsars, and they could be better by as much as an order of magnitude if the masses and radii were known sufficiently accurately.   

The study presented here is by no means intended as a final word on tests of GR with \LISA observations of GWs emitted by galactic binaries; rather, it hopefully opens the door to future studies of the extraction of theoretical physics from a guaranteed \LISA source that has not received much attention in the GW experimental relativity community. Both the model-independent and the model-dependent frameworks ought to be more thoroughly investigated to account for systematics that could arise from astrophysical effects, like mass transfer and winds. The latter should also be extended to other modified theories of gravity that also predict the existence of pre-Newtonian dissipation effects. Nonetheless, the work presented here will hopefully act as a igniter that will stimulate such future studies.    

The remainder of this paper presents the details of the calculations that lead to the results described above. 
Section~\ref{sec:WDWD-modeling} describes how to model WD binaries outside of GR. 
Section~\ref{sec:Data-Analysis} discusses the data analysis techniques we will use to determine the magnitude of projected constraints with \LISA. 
Section~\ref{sec:model-independent-constraints} presents the projected constraints we can place on model-independent deformations of the GR model prediction. 
Section~\ref{sec:model-dependent-constraints} focuses on scalar-tensor theories and presents model-dependent constraints. 
Section~\ref{sec:caveats} briefly discusses caveats that could modify the conclusions presented earlier quantitatively but not qualitatively. 
Section~\ref{sec:conclusions} summarizes the main results of this work and concludes by discussing possible future work.  
Throughout this paper, we use the following conventions: $G = c = 1$, $(\mu, \nu, \rho, ...)$ are spacetime indices.

\section{Galactic Binary Modeling with pre-Newtonian Dissipation}
\label{sec:WDWD-modeling}

This section describes how to model GWs emitted by galactic binaries appropriate for \LISA science. We then conclude with a discussion of astrophysical populations of galactic binaries 

\begin{table*}[htp]
\begin{center}
\begin{tabular}{ c | c c c c c c c c c | c c}
\toprule
\multicolumn{12}{c}{WD-WD Parameters} \\
Source & $\f$ (s$^{-1}$)& $\fdot$ (s$^{-2}$) & $\fddot$ (s$^{-3}$)& $\mathcal{A}$ & $\cos\theta$ & $\phi$ (rad)& $\cos\iota$ & $\psi_{\rm GW}$ (rad)& $\varphi_0$ (rad) & $\Mc$ ($\Msun$) & \SNR (1 yr)\\
\hline
0 & 0.015248 & 2.5834e-14 & 1.6049e-25 & 7.7243e-23 & -0.0708 & 4.6600 & -0.2016 & 0.0430 & 4.2766 & 0.39 & 61 \\ 
1 & 0.008994 & 1.7922e-14 & 1.3094e-25 & 1.4888e-22 & 0.2779 & 4.8717 & 0.5001 & 1.2662 & 6.1245 & 0.99 & 186 \\ 
2 & 0.012124 & 1.6886e-14 & 8.6230e-26 & 7.8345e-23 & -0.5828 & 4.3243 & 0.5973 & 0.1514 & 3.1676 & 0.50 & 108 \\ 
3 & 0.012163 & 1.5602e-14 & 7.3380e-26 & 1.0910e-22 & 0.0909 & 4.7667 & 0.2371 & 1.9003 & 2.5800 & 0.47 & 97 \\ 
4 & 0.018509 & 4.4578e-14 & 3.9367e-25 & 6.4157e-23 & -0.5394 & 4.3525 & 0.7049 & 0.4413 & 2.2722 & 0.35 & 95 \\ 
5 & 0.007984 & 7.9289e-15 & 2.8873e-26 & 4.7292e-22 & -0.7201 & 4.0816 & -0.4880 & 0.8372 & 2.2434 & 0.79 & 560 \\ 
6 & 0.015234 & 1.3389e-14 & 4.3146e-26 & 8.3676e-23 & 0.0719 & 4.6291 & 0.2790 & 2.1810 & 4.3866 & 0.26 & 77 \\ 
7 & 0.015144 & 2.0886e-14 & 1.0562e-25 & 1.0816e-22 & -0.7246 & 3.8508 & 0.4231 & 2.0689 & 3.1987 & 0.35 & 113 \\ 
8 & 0.017695 & 1.9882e-13 & 8.1911e-24 & 3.6914e-22 & 0.7288 & 5.4111 & -0.3325 & 1.9704 & 3.6658 & 0.95 & 329 \\ 
9 & 0.012800 & 1.5497e-14 & 6.8793e-26 & 7.5127e-23 & -0.3924 & 4.4753 & 0.4679 & 1.3264 & 0.4712 & 0.42 & 89 \\ 
\hline
\multicolumn{12}{c}{} \\
\multicolumn{12}{c}{WD-NS Parameters} \\
Source & $\f$ (s$^{-1}$)& $\fdot$ (s$^{-2}$) & $\fddot$ (s$^{-3}$)& $\mathcal{A}$ & $\cos\theta$ & $\phi$ (rad)& $\cos\iota$ & $\psi_{\rm GW}$ (rad)& $\varphi_0$ (rad) & $\Mc$ ($\Msun$) & \SNR (1 yr)\\
\hline
$1.4+0.3\Msun$ & 0.003000 &1.1422e-16  & 1.5945e-29  & 5.4222e-23 &  -0.1494 & 4.600 & 0 & 0 & 0 &  0.53 & 36 \\
-- & 0.005000 & 7.4332e-16  & 4.0519e-28   & 7.6221e-23 & -- & -- & -- & -- & -- & -- & 88 \\
-- & 0.010000 & 9.4396e-15  & 3.2672e-26  & 1.2099e-22 & -- & -- & -- & -- & -- & -- & 91 \\
-- & 0.020000 & 1.1988e-13  & 2.6345e-24 & 1.9207e-22 & -- & -- & -- & -- & -- & -- & 214 \\
$1.4+1.3\Msun$ & 0.010000 & 3.5059e-14  & 4.5069e-25  & 4.4938e-22 & -- & -- & -- & -- & -- & 1.17 & 753 \\
\end{tabular}
\end{center}
\caption{Parameters of the ten WD-WD systems and the five WD-NS systems considered in our analysis. 
The parameters $(\f,\fdot,\fddot)$ describe the orbital evolution, while $\mathcal{A}$ is the dimensionless gravitational wave amplitude. 
The sky location is encoded in $(\cos\theta,\phi)$ where $\theta$ is the co-latitude and $\phi$ is the longitude in ecliptic coordinates.
The angular parameters that encode the binary's orientation are ($\cos\iota,\psi_{\rm GW},\varphi_0$) where $\iota$ is the inclination angle, $\psi_{\rm GW}$ is the polarization angle, and $\varphi_0$ is the orbital phase at $t=0$.
Quantities derived from the source parameters are $\Mc$ and the signal to noise ratio $\SNR$ of the binary after one year of observing (which will grow roughly as $\sqrt{T}$).
Dashees indicate the parameter value is identical to the row above.
All sources are simulated to be consistent with GR, so $\fddot$ can be derived from $\f$ and $\fdot$.
The WDWD systems are drawn from a spatial distribution consistent with the Milky Way galaxy, with random orientations uniform on a sphere.
The WDNS systems are chosen to be edge-on (therefore eclipsing, to maximize what can be learned from EM observations) simulated with $\mathcal{A}$ to be consistent with the source at a distance of 8 kpc.
}
\label{tbl:wdwd}
\end{table*}%

\subsection{Modeling Individual Sources}
\label{subsec:generic-modeling}

%
%
%
%
%
%
%
%
%

We model the \LISA response to GWs from galactic binaries using the fast-slow decomposition described in~\cite{Cornish:2007}.
The waveform model is extended to include the second time derivative of the frequency, modifying the GW phase $\Psi(t) = 2 \pi \f t + \pi \fdot t^2 + \frac{1}{2}  \pi \fddot^3 + \varphi_0$ where $\f$ and $\varphi_0$ are the GW frequency and phase measured at some fiducial time, such as when observations begin, and the over dots denote time derivatives evaluated at the fiducial time.

In the waveform model, the parameters $\vec{\lambda} = \{\f,\fdot, \fddot,\ldots \}$ are independent, but in any given theory of gravity, there are inter-relations between them. For example, in GR, we have that
\begin{align}
\label{eq:fdotGR}
\fdot_{\GR} &= \frac{96}{5 \pi} {\cal{M}}^{-2} \left(\pi {\cal{M}} \f\right)^{11/3}\,,
\\
\label{eq:fdotGR2}
\fddot_{\GR} &= \frac{33792}{25 \pi} {\cal{M}}^{-3} \left(\pi {\cal{M}} \f\right)^{19/3}\,,
\end{align}
to leading order in an expansion about ${\cal{M}} f \ll 1$, where ${\cal{M}} = \eta^{3/5} m$ is the chirp mass, $\eta = m_{1} m_{2}/m^{2}$ is the symmetric mass ratio and $m = m_{1} + m
_{2}$ is the total mass. Therefore, assuming the system dynamics are dominated by GW emission, and assuming the null hypothesis that GR is correct, a measurement of $(\f,\fdot)$ suffices to extract the chirp mass ${\cal{M}}$. The measurement of any higher derivative can then serve as a consistency check of the null hypothesis.  

When pre-Newtonian dissipation effects are present, the relation between the components of $\vec{\lambda}$ changes. Generically, one can show that the chirping rate now becomes~\cite{Yunes:2009ke}
\begin{align}
\label{eq:fdot-mapping}
\fdot &= \fdot_{\GR} \left[1 + \beta_{\ppE}  \left(\pi {\cal{M}} f\right)^{b_{\ppE}}\right]\,,
\end{align}
which then implies that
\begin{align}
\label{eq:fddot-mapping}
\fddot &= \fddot_{\GR} \left[1 
+ \frac{20}{11} \beta_{\ppE}  \left(\pi {\cal{M}} f\right)^{b_{\ppE}}  + {\cal{O}}\left( \beta_{\ppE}^{2} (\pi {\cal{M}} f)^{2 b_{\ppE}}\right) \right],
\end{align}
where $\beta_{\ppE} \in \Re$ and $b_{\ppE} \in \Re$ are parameterized post-Einsteinian amplitude and exponent parameters~\cite{Yunes:2009ke}. We have here neglected terms proportional to $ \beta_{\ppE}^{2} (\pi {\cal{M}} f)^{2 b_{\ppE}}$ since we assume that $\beta_{\ppE} \ll (\pi {\cal{M}} f)^{-b_{\ppE}}$.  The amplitude parameter $\beta_{\ppE}$ can in principle depend on other dimensionless system parameters, like the symmetric mass ratio. The exponent coefficient $b_{\ppE} < 0$ in order for the effect to be pre-Newtonian, with $b_{\ppE} = -2/3$ representing the dipole effect we will focus on in this paper.   

\subsection{Astrophysical Population}

Of the tens of thousands of galactic binaries expected to be detectable in the \LISA band, population synthesis simulations suggest that several such systems will be exceptionally strong sources. These binaries will be detectable within the first few weeks of mission operations, and over years of observing will accumulate signal to noise ratios (\SNR) in the hundreds. For testing GR with galactic binaries, many of the loudest \LISA sources are the best laboratories not only because of their high \SNR, but also because these tend to be higher-mass WD binaries (with $\Mc\sim0.4\, \text{M}_\odot$, corresponding to component masses of $\sim0.5\, \text{M}_\odot$ for equal-mass binaries). As a consequence, the individual stars will be more compact, suppressing contributions to the orbital evolution of the binary through gravitational interactions between the stars (e.g., tidal interactions or mass transfer). 

For our investigations we will select ``loud,'' non-interacting binaries from a galaxy realization consistent with those used for the \LISA Data Challenges\footnote{\url{https://gitlab.in2p3.fr/stas/MLDC}} constructed from simulations by~\cite{Toonen:2012jj}. 
From the galaxy simulation, binaries are ranked by \SNR\ and the 10 loudest systems with $\f>4$ mHz are selected for study. 
There are high \SNR\ binaries at lower frequencies, but they will more likely be contending with overlapping signals, time-varying galactic confusion noise, and, for lower mass systems, dynamical influences other than gravity due to mass transfer, tidal interactions, etc.  
In general, exploiting galactic binaries for testing orbital dynamics will be most straightforward using isolated systems at high frequency. The binary parameters for the systems we will study are presented in Table \ref{tbl:wdwd}.

In addition to the astrophysical population of WD binaries, we will also investigate what can be achieved with \LISA should we observe systems composed of a WD and a NS. In our simulations, the NS-WD binaries are placed near the galactic center (where the source density of galactic binaries is likely the highest), and we study five canonical systems, varying the observing time, masses,  and the orbital frequencies of the binary. The binary parameters for these systems are also included in Table~\ref{tbl:wdwd}.

\section{Data Analysis}
\label{sec:Data-Analysis}

\subsection{LISA}
LISA will be a constellation of three free-flying spacecraft, each on independent heliocentric orbits approximately one AU from the Sun.
Each satellite is equipped with an optical metrology system to precisely monitor the distances between free-falling test masses housed within the spacecraft.
From the inter-spacecraft ranging measurements, two Michelson-like interferometry signals are synthesized digitally to simultaneously measure the two GW polarizations, while a third Sagnac channel is insensitive to GWs (at low frequencies) and will be used for detector characterization.
The simulated detector performance used in this study is consistent with the instrument noise levels quoted in the \LISA L3 Mission Proposal~\cite{Audley:2017drz}.
This analysis takes advantage of the noise-orthogonal $A$ and $E$ data streams constructed from linear combinations of the Michelson-like time delay interferometry channels~\cite{Tinto:2002de}.
The $A$ and $E$ data streams assume identical noise in each of the six interferometer links of the spacecraft, and so are an idealized case that will not be realized during mission operations, but are suitable for this proof-of-concept study.

\subsection{Methodology (Bayes/MCMC)}
The simulated galactic binary sources are analyzed using a parallel tempered Markov Chain Monte Carlo (MCMC) pipeline adapted from~\cite{Littenberg:2011zg} specifically designed for galactic binary searches in \LISA data.
The MCMC pipeline has been modified to include the second time derivative of the frequency as a parameter.
We assume uniform priors on the frequency parameters. Mapping the gravitational-wave observables to the modified theory parameters is done through post-processing of the posterior samples from the Markov chain, the specifics of which will be described in later sections.

\subsection{Example Parameter Estimation Results}
\label{subsec:example}
In this work we will be mapping the parameters used by the MCMC sampler to some other physical quantities and evaluating the posterior distribution function at arbitrary parameter values.  
The latter is generally a challenging problem as the MCMC yields a discrete set of samples drawn from the distribution, rather than the continuous distribution itself.  
Here we are spared much of this difficulty because our investigations involve only extraordinarily high \SNR sources, and the high \SNR limit is where the errors are well approximated by Gaussians distributions.

\begin{figure}[htb]
	\includegraphics[width=0.5\textwidth]{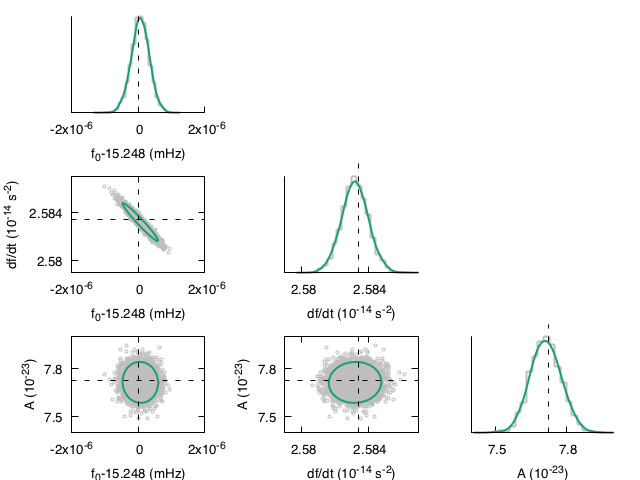}
	\caption{(Color online) Corner plot showing binned marginalized posteriors (gray histograms) of model parameters along the diagonal, and scatter plot (gray dots) of MCMC samples for the different two-dimensional combinations in the off diagonal cells. Parameters, from top left to bottom right along the diagonal, are the initial GW frequency $\f$, first time derivative of the frequency $\fdot$, and GW amplitude $\mathcal{A}$. We approximate the joint posteriors by a multivariate Gaussian, with covariance matrix computed from the MCMC samples. Colored curves along the diagonals show the Gaussian fit from the approximation, and the $2\sigma$ contours are shown in the off-diagonal plots. These results validate our approximation of the continuous posterior distribution function using a multivariate Gaussian fit to the MCMC samples. Dashed lines indicate the true source parameters of the simulation.}
	\label{fig:PE_frequencies.png}
\end{figure}

To demonstrate this, we show 1D histograms and 2D scatter plots of the samples from the MCMC for a standard analysis of Source 0 (i.e., not including $\fddot$ as a parameter), assuming a three year observation time.
The parameters shown in Fig.~\ref{fig:PE_frequencies.png} are the GW amplitude  $\mathcal{A}$, the GW frequency $\f$, and its first time derivative $\fdot$.  
The posterior distribution function $p(\vec{\lambda}\vert d)$ is then approximated by a multivariate Gaussian 
\be
p(\vec{\lambda}\vert d) \approx \frac{1}{(2\pi \bf\Sigma)^{1/2}} e^{-\frac{1}{2}(\vec\lambda-\vec\mu)^T\bf\Sigma^{-1}(\vec\lambda-\vec\mu)}
\label{eq:multivariate-gaussian}
\ee
centered at the mean of the parameters $\vec\mu$ with covariance matrix ${\bf \Sigma}$ calculated directly from the chain samples, $\Sigma_{ij} = \langle \lambda_i \lambda_j \rangle$.
The green curves in Fig.~\ref{fig:PE_frequencies.png} show the Gaussian representation of the 1D posteriors and the $2\sigma$ contours of the covariance matrix computed from the MCMC samples.

\begin{figure}[htb]
	\includegraphics[width=0.5\textwidth]{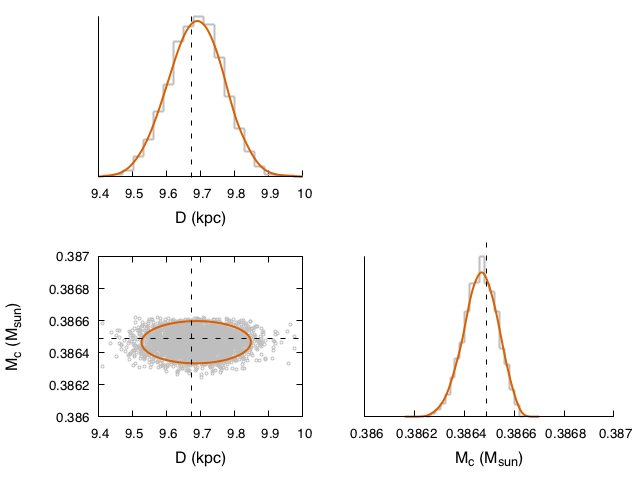}
	\caption{(Color online) Same as Fig.~\ref{fig:PE_frequencies.png}, but now showing the chirp mass $\Mc$ and luminosity distance $D_L$ derived from the frequency and amplitude parameters of the model, assuming no detectable contribution to the orbital evolution from astrophysical or non-GR effects. Dashed lines indicate the true source parameters of the simulation.}
	\label{fig:PE_derived.png}
\end{figure}

Figure~\ref{fig:PE_derived.png} shows an example of how the MCMC samples can be mapped to different physical quantities. In particular, we here show the canonical example of mapping $(\mathcal{A}, \f, \fdot)$ to the luminosity distance and chirp mass $(D_L,\Mc)$. At every value of $(\mathcal{A}, \f, \fdot)$ sampled by the chains, we compute the corresponding $(D_L,\Mc)$, and then construct a histogram of the result (gray in the figure). We then use the Gaussian approximation to approximate the posterior distributions for $(D_L,\Mc)$ using the covariance matrix of the reparameterized samples (orange in the figure). Once more, observe how the Gaussian estimate is an excellent approximation to the sampled posteriors.  Beyond confirming by eye the agreement, we test that the p-values of the MCMC samples evaluated with the approximate posterior are consistent with being uniformly distributed between $[0,1]$.

The most important parameter for our study is the second time derivative of the frequency, $\fddot$, as later in this work it will either be used to place an independent constraint on the chirp mass, or to break degeneracies with the pre-Newtonian dipole term.
The $\fddot$ parameter is often neglected in GW studies of WD binaries because it is seldom measurable. Moreover, this parameter provides redundant (though less constraining) information about $\Mc$ and $D_L$, which are the two parameters typically most sought after when thinking about exploiting the astrophsical predictions of the frequency evolution in galactic binaries.

\begin{figure}
	\includegraphics[width=0.5\textwidth]{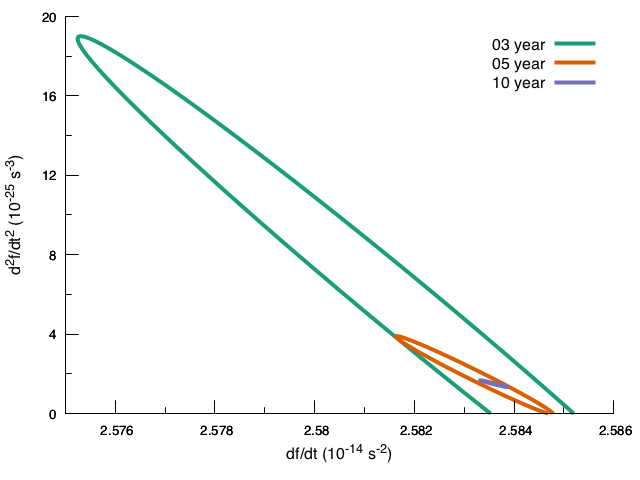}
	\caption{(Color online) $2\sigma$ contours from a multivariate Gaussian approximation using the covariance matrix of the MCMC samples marginalized to the $\fdot\text{--}\fddot$ plane. Different ellipses show how the measurement of the frequency evolution improves over the observing time. For this source, a 5 year observation was needed to detect a non-zero $\fddot$, and a 10 year observation is required for a precise constraint. This conclusion is representative of the $\fddot$ measurement for the other simulated WD-WD binaries we studied.}
	\label{fig:fdot_fddot.png}
\end{figure}

For this work, the story of how well WD binaries can be used to test gravity is the story of how well $\fddot$ can be measured. To that end, we focus on \emph{golden galactic binaries}, the best-of-the-best in a simulated galactic population: high mass, short orbital period, systems which will undergo the most significant frequency evolution during the \LISA observations, and will be easily detectable anywhere in the galaxy.
Even with these extraordinary sources, long integration times are needed for the observations to constrain $\fddot$.
Figure~\ref{fig:fdot_fddot.png} shows the $2\sigma$ contours in the $\fdot\text{--}\fddot$ plane from posterior distribution functions represented by the MCMCs covariance matrix.
On display here is the same source as shown in Figs.~\ref{fig:PE_frequencies.png} and ~\ref{fig:PE_derived.png}, now including $\fddot$ as one of the model parameters.  
This figure demonstrates how $\fddot$ improves with observing time.  
During the first 3 years of observing the second derivative of $\f$ is basically unmeasured.  
After 5 years there are hints of a detection, as the parameter is constrained to be non-zero.
For this simulated source, a 10 year observation is required to unambiguously measure a non-zero $\fddot$.
	
\section{Projected Model-independent Constraints}
\label{sec:model-independent-constraints}

In this section we present the details of the model-independent test. We begin by describing the injection and the template that we used. We then proceed with the details of the test and present results from our data analysis studies. 

\subsection{GR injection and dipole template}

\begin{figure*}
	\includegraphics[width=1.0\textwidth]{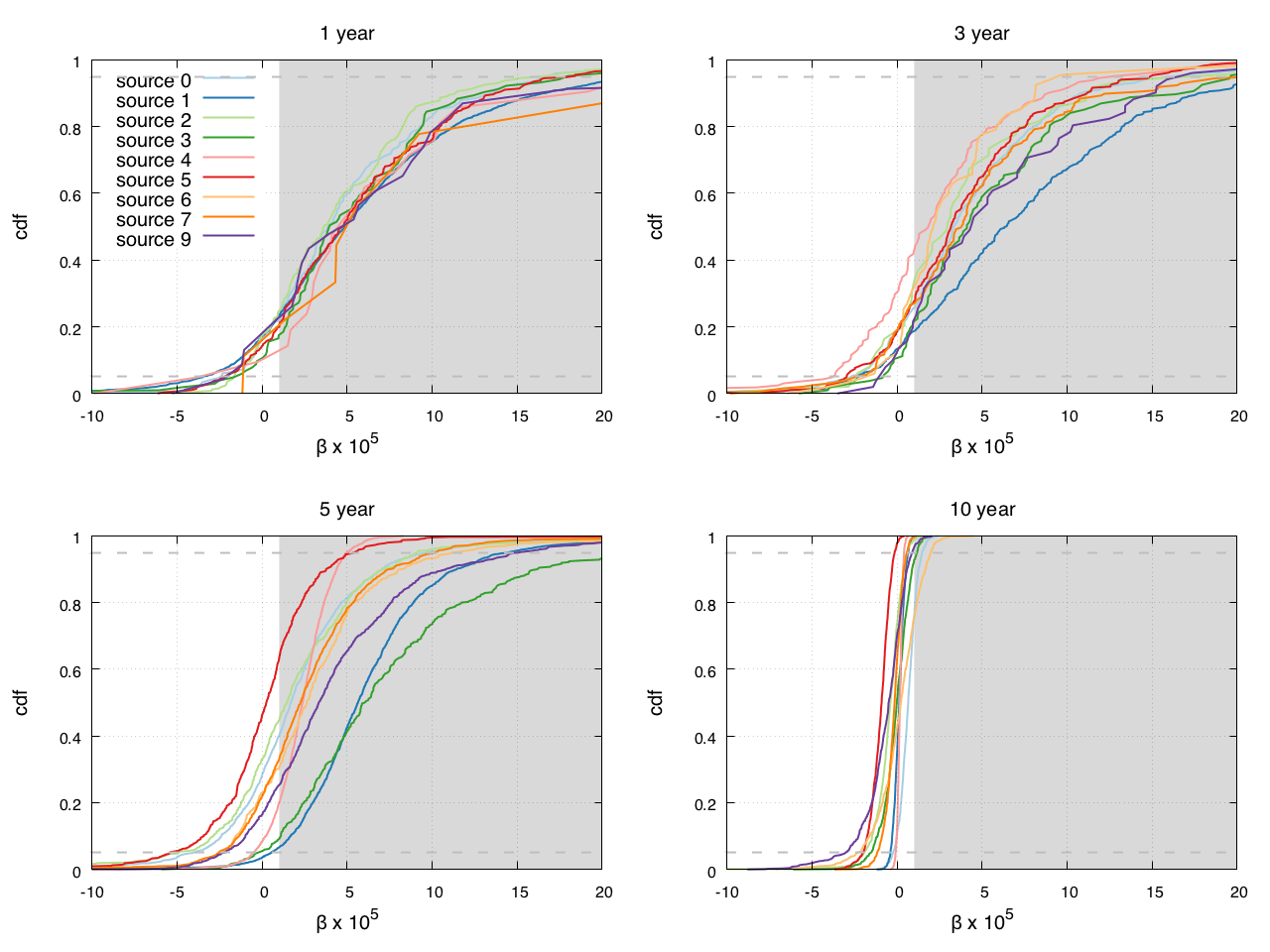}
	\caption{(Color online) Cumulative distribution function of the marginalized posterior distribution function for $\beta$ from WD-WD binaries in our study as a function of observing time. The binaries are simulated to be consistent with GR ($\beta=0$). The gray shaded region is where our approximation that dipole radiation is a small perturbation to the GW emission breaks down. The measurements of $\beta$ are dominated by uncertainty in the determination of $\fddot$. Long observing times are required to put tight constraints on $\beta$. }
	\label{fig:beta_cdf_only.png}
\end{figure*}

As a first demonstration of using WD binaries to test GR we take a model-agnostic approach, instead constraining the magnitude of the dipole effect on the waveforms.
To do so, we analyze the ten binaries selected from the simulated galaxy as the most promising sources for these tests using the MCMC pipeline with $\fddot$ included as a parameter.
The binaries are simulated with $\beta_{\rm ppE}=0$, i.e. consistent with GR, and are then recovered using $\f,\fdot$ and $\fddot$ as independent parameters, instead of being constrained to follow the GR relationship.
For the remainder of this section, we will drop the ppE subscript on $\beta$ for convenience.
The marginalized posterior $p(\f,\fdot,\fddot \vert d)$ is then mapped to $p(\Mc,\beta\vert d)$ by inverting Eqs. \eqref{eq:fdot-mapping} and \eqref{eq:fddot-mapping}.  

Figure~\ref{fig:beta_cdf_only.png} shows cumulative distribution functions of the marginalized posterior on $\beta$, $p(\beta\vert d)$ for each source as a function of observation length.
Horizontal dashed lines denote the $5\%$ and $95\%$ percentiles, enclosing the $90\%$ credible interval of the measurement. 
Recall that the $\beta$ parameter can in principle depend on the symmetric mass ratio $\eta$, and therefore, if it does, then $\beta$ is not the same for all binaries in our study.
As a result, we are not combining results to establish a joint posterior on $\beta$, instead treating the $\beta$ measurement of each source independently.
The gray shaded region at $\beta\gtrsim10^{-5}$ is roughly where the approximation in Eqs. \eqref{eq:fdot-mapping} and \eqref{eq:fddot-mapping} breaks down, with $\beta(\pi\Mc\f)^{-2/3}\gtrsim 1$. 
The exact place where the $\beta$ correction becomes too large depends on the system's mass and orbital period, which is comparable for all of the binaries we consider here, but not identical. 

From these results we see that uncertainties in the measurement of $\beta$ are typically of order $10^{-4}$ for observation times of a few years, improving to ${\lesssim}10^{-5}$ after monitoring the sources for ten years.
For the binaries to be measured well enough to definitively constrain $\beta$ within the region where the ppE formalism is valid, ${\sim}$decade long monitoring campaigns are needed. 

\subsection{Redundancy Test and Fundamental Bias}

Let us now consider the effect on WD binary parameter estimation if we assume GR is correct in the analysis, but include the possibility of non-zero dipole radiation in the injected source simulations.
Treating each marginalized posterior of the orbit parameters $p(\f\vert d),p(\fdot\vert d),p(\fddot\vert d)$ independently, we can compute three independent constraints on the frequency-dependent chirp mass function $\Mc(f_{0})$, derived from the inversion of Eq.~\eqref{eq:fdotGR}, or Eqs.~\eqref{eq:fdot-mapping} and~\eqref{eq:fddot-mapping}.
Doing so ignores the covariances between the different frequency parameters, which are shown to be non-negligible in Figs.~\ref{fig:PE_frequencies.png} and ~\ref{fig:fdot_fddot.png}, leading to a conservative estimate.

\begin{figure*}
	\includegraphics[width=0.49\textwidth]{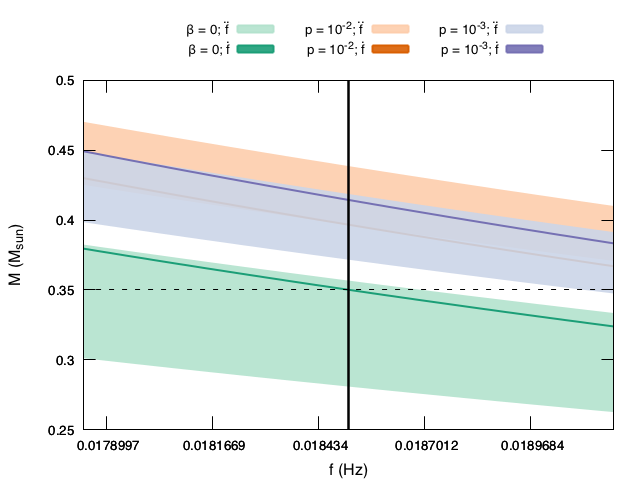}
	\includegraphics[width=0.49\textwidth]{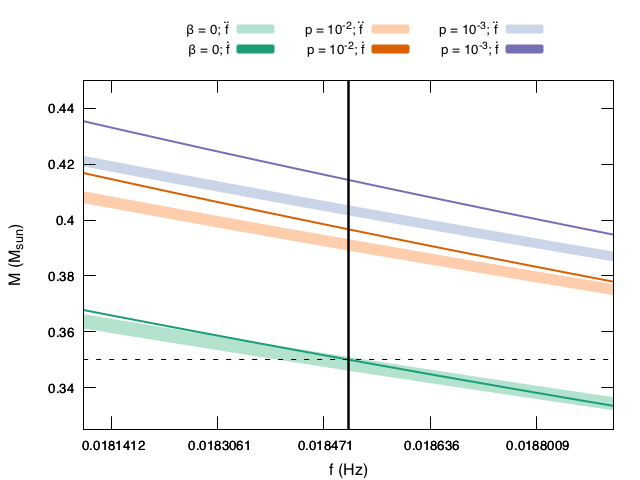}\;
	\caption{(Color online)  Left: Consistency plots for GR injection like in Fig.~\ref{fig:Mc_f.png} for a 5 year observing time.  The WD-WD systems are simulated to be consistent with GR (green), and with increasingly large $\beta$, injected at the 0.99 (orange) and 0.999 (purple) quantiles of the 10 year CDF from Fig.~\ref{fig:beta_cdf_only.png}. Note that for each source the frequency parameters yield a self-consistent value of the chirp mass, but only the $\beta=0$ source recovers the true value (horizontal, dashed line). Right: Same as the left figure, but after a 10 year observing time. Now the frequency parameters are measured precisely enough that the consistency test is failed for the non-GR simulations.}
	\label{fig:Mc_f_beta0.png}
\end{figure*}

If the sources' frequency evolution are consistent with GR, the three constraints will intersect in $\Mc-\f$ space.
Figs.~\ref{fig:Mc_f.png} and ~\ref{fig:Mc_f_beta0.png} show $90\%$ credible intervals for $\Mc(f)$ from two of the detached WD binaries used in this study.
The vertical line is the constraint from $p(\f\vert d)$, which spans the prior range on $\Mc$.
The bands sweeping from top left to bottom right are the constraints from $p(\fdot\vert d)$ (thin) and $p(\fddot\vert d)$ (thick).
The horizontal dashed line indicates the true value of the chirp mass from the simulated signal.

Similar to Fig.~\ref{fig:Mc_f.png}, Fig.~\ref{fig:Mc_f_beta0.png} shows constraints for different injected sources with a five year (left panel) and ten year (right panel) observing time. When the injected source is simulated with a GR frequency evolution (green bands), one obtains a redundancy test like the one shown also in Fig.~\ref{fig:Mc_f.png}. Observe again that as the observing time increases, the redundancy test becomes stronger. When the injected source is simulated with a non-GR frequency evolution (orange and purple bands), one sees how the failure of the redundancy test could indicate a departure from GR.  

The non-GR detections were simulated as follows. We used the frequency evolution in Eqs.~\eqref{eq:fdot-mapping} and~\eqref{eq:fddot-mapping} with $\beta \neq 0$, but still within the regime where $\beta(\pi\Mc\f)^{-2/3} \ll 1$.
We use Source 4 as the test candidate, as it yielded some of the best constraints on $\beta$. 
The non-zero $\beta$'s were selected to be at the 0.99 and 0.999 percentile of the $p(\beta|d)$ from the 10 year observation in Fig.~\ref{fig:beta_cdf_only.png}.
The orange bands show results for the injection with $p$-value of $10^{-2}$, while the purple bands are for the source with non-zero $\beta$ at $p$-value of $10^{-3}$.

This demonstration leads to two notable conclusions.  
First, the left panel of Fig.~\ref{fig:Mc_f_beta0.png}, coming from a simulated 5 year observation, shows that the uncertainties in the $\fddot$ measurement are such that signals whose underlying evolution is \emph{not} consistent with GR can still pass the $\Mc$ consistency tests insofar as the three different constraints intersect in the same region of $\Mc-f$ space.
However, the altered dynamics lead to a systematic or \emph{stealth} bias~\cite{Yunes:2009ke,Vallisneri:2013rc} in the inferred chirp mass that is larger than the statistical error (roughly the width of the intersection between the $p(\f\vert d)$ and $p(\fdot\vert d)$ constraints).
In other words, beyond-relativistic frequency evolution can meaningfully affect our inferences about the source before it is large enough to be directly detected.
Without an accurate, independent measurement of $\Mc$ from EM follow-up observations, this effect would go unnoticed.

The right panel of Fig.~\ref{fig:Mc_f_beta0.png} panel shows how this \emph{stealth bias} is exposed by continued monitoring of the source, showing the extreme case of a 10 year observation.
Here the constraints on the frequency evolution are tight enough that the bands are clearly not intersecting, proving that the assumption of only GR-driven evolution is not supported by the data.

\begin{figure*}
	\includegraphics[width=0.49\textwidth]{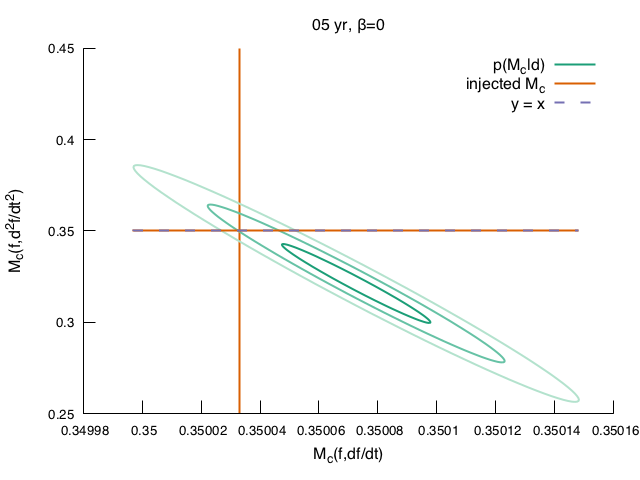} \;
	\includegraphics[width=0.49\textwidth]{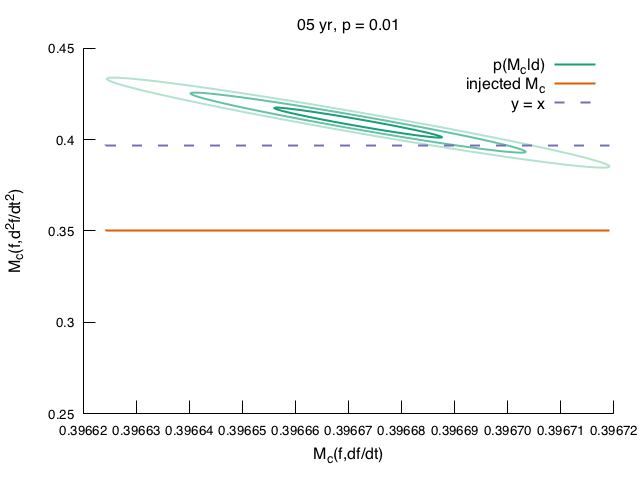}
	\caption{(Color online)  Left: 1,2, and 3$\sigma$ contours of the multivariate gaussian approximation to the marginalized posterior of the chirp mass using $\Mc(\f,\fdot)$ (horizontal axis) and $\Mc(\f,\fdot)$ (vertical axis) for a GR injection ($\beta=0$). The orange solid lines denote the true value of $\Mc$ for the simulated source, the purple dashed line denotes the line $y=x$ where the two $\Mc$ measurements are self-consistent.  Right: Same figure but now the true source has a non-zero $\beta$, chosen to be at the $0.99$ quantile of the 10 year CDF from Fig.~\ref{fig:beta_cdf_only.png}. The posterior intersects the line $y=x$ and therefore supports a self-consistent measurement of the $\Mc$, but confidently \emph{excludes} the true value by assuming zero dipole radiation. Both results are for 5 year observing times.}
	\label{fig:Mc_Mc_p010.png}
\end{figure*}

The notion of non-zero $\beta$ fundamentally biasing chirp mass measurements is further illustrated in Fig.~\ref{fig:Mc_Mc_p010.png}, now showing the 2D marginalized posterior $p(\Mc(\f,\fdot),\Mc(\f,\fddot) \vert d)$ (and therefore taking advantage of the information encoded in the parameter correlations).
Here the covariance matrix of the MCMC samples has been used to approximate the posterior as a multivariate Gaussian, and we are plotting the 1, 2, and 3$\sigma$ contours in green.
The line $y=x$ (purple, dashed) denotes where the two chirp mass constraints are self-consistent, i.e. $\Mc(\f,\fdot)$ and $\Mc(\f,\fddot)$ return the same value. 
The solid orange lines mark the true value of the simulated source.
The left panel is from a 5 year observation of a source with $\beta=0$, and we find that the posterior encompasses $y=x$ and encloses the true value.

The right hand panel shows results from the same duration observation and same source, but with $\beta$ injected at the $p$-value of $10^{-2}$ of the 10 year constraint.
Note that the posterior includes the line $y=x$ so, without independent information about the true $\Mc$, this source would appear to be consistent with GR-only frequency evolution. 
However, the true value (orange line) is several $\sigma$ away from the median of the distribution, resulting in a ${\sim}15\%$ bias in the inferred $\Mc$ (and closer to $30\%$ for the $p\sim10^{-3}$ case, not shown here).

\section{Projected Model-Dependent Constraints}
\label{sec:model-dependent-constraints}

In this section we specialize the model-independent test of the previous section to scalar-tensor theories. We begin by describing the basics of this theory and establishing some notation. We proceed by presenting projected constraints assuming GW observations can be used to extract $(\f,\dot{f},\ddot{f})$ in connection with electromagnetic observations of the same source, which can provide the masses and radii of the binary components. Throughout this section, we mostly follow the notation of~\cite{Damour:1998jk} for scalar-tensor theories. 

\subsection{Basics of Scalar-Tensor Theories and the Modeling of Individual Sources}

The class of scalar-tensor theories we will consider can be defined by the Einstein-frame action
\begin{align}
S &= \frac{1}{2 \kappa_{*}} \int d^{4}x  \sqrt{-g_{*}} \left(R_{*}  - 2 g^{\mu \nu}_{*} \partial_{\mu} \psi \partial_{\nu} \psi \right) \\
&+ S_{M}\left[\chi_{M}, A^{2}(\psi) g_{\mu \nu}^{*}\right]\,,
\end{align}
where $\kappa_{*} = 8 \pi G_{*}$, with $G_{*}$ the bare gravitational constant in the Einstein frame, $\psi$ is a massless scalar field in the Einstein frame (not to be confused with the gravitational wave polarization angle $\psi_{\rm GW}$ from above), $R_{*}$ is the Ricci scalar in the Einstein frame associated with the Einstein frame metric $g_{\mu \nu}^{*}$, with $g_{*}$ its determinant, $\chi_{M}$ are additional matter degrees of freedom and $A(\psi)$ is a coupling function. 

The action above describes a theory in which matter follows geodesics of the physical (Jordan-frame) metric
\be
g_{\mu \nu} = A^{2}(\psi) g_{\mu \nu}^{*}\,.
\ee
We are here interested in a particular subclass of scalar-tensor theories defined by the conformal coupling
\be\label{eqphi}
A(\psi) = \exp\left[\alpha_{0} \left(\psi - \psi_{0}\right) + \frac{1}{2} \beta_{0} \left(\psi - \psi_{0}\right)^{2}\right]\,,
\ee
where $\psi_{0}$ is the ``vacuum-expectation value'' of the field at spatial infinity. One could conformally transform the Einstein-frame action to the Jordan frame, but the resulting action is more difficult to work with, and so, we here choose to present equations in the Einstein frame only. The physical world in which experiments are performed, however, exists in the Jordan frame, so one must always remember to re-cast any Einstein frame results to the Jordan frame before comparing against experiments.

The field equations in the Einstein frame take the form
\begin{align}\label{mod_eins}
G_{\mu \nu}^{*} &= \kappa_{*} T_{*}^{M}{\!\!\!}_{\mu \nu} + T_{*}^{\psi}{\!\!}_{\mu \nu} \,,
\\
\label{eq:scalar-field-eq}
\square_{*} \psi &=- \frac{\kappa_{*}}{2} \left[ \alpha_{0} + \beta_{0} \; \left(\psi - \psi_{0}\right)\right] \; T^{M}_{*}\,, 
\end{align}
where $G_{\mu \nu}^{*}$ is the Einstein tensor in the Einstein frame, $T_{*}^{M}{\!\!\!}_{\mu \nu}$ is the matter stress-energy tensor in the Einstein frame, and the stress-energy tensor for the Einstein-frame scalar field is
\begin{align}
T_{*}^{\psi}{\!\!}_{\mu \nu} &= 2 \partial_{\mu} \psi \partial_{\nu} \psi -  g_{\mu \nu}^{*} \; \left(g^{\alpha \beta}_{*} \partial_{\alpha} \psi \partial_{\beta} \psi\right)\,,
\end{align}
and $T^{M}_{*}$ is the Einstein-frame trace of the matter stress-energy tensor $T_{*}^{M}{\!\!\!}_{\mu \nu}$ in the Einstein frame. Taking the divergence of the field equations for the metric, one finds an additional compatibility condition, which guarantees the physical metric satisfies the weak-equivalence principle. 

In the Solar System, this theory can reduce closely to GR. To show this, let us say the Einstein frame field $\psi \to \psi_{0}$ in the Solar System. Then, at 1 PN order in the Solar System, Damour and Esposito-Far\'ese have shown that the parametrized post-Newtonian (ppN) parameters~\cite{Nordtvedt:1968qs,Will:1972zz} take the form~\cite{Damour:1992we,Damour:1993hw,Damour:1996ke,Damour:1998jk} 
\begin{align}
\left|\gamma_{\ppN} - 1 \right| &= 2 \frac{\alpha_{0}^{2}}{1 + \alpha_{0}^{2}}\,,
 \\
\left| \beta_{\ppN} - 1 \right| & = \frac{1}{2} \frac{\beta_{0} \alpha_{0}^{2}}{(1 + \alpha_{0}^{2})^{2}}\,.
\end{align}
The observation of the Shapiro time delay through tracking of the Cassini spacecraft can be satisfied by requiring that $\alpha_{0}^{2} \lesssim 10^{-3}$, without imposing any condition on $\beta_{0}$~\cite{will-living,Bertotti:2003rm}. 

Even with a small value of $\alpha_{0}$, large modifications to GR can still occur in systems that are strongly gravitating. Damour and Esposito-Far\'ese have shown that an effect akin to ferromagnetism occurs in the vicinity of sufficiently compact and massive stars, which leads to a non-linear activation of the scalar field, an effect called \emph{spontaneous scalarization}~\cite{Damour:1992we,Damour:1993hw,Damour:1996ke,Damour:1998jk}. In numerical simulations, such scalarization has also been seen to occur dynamically in NS binaries when the binary's binding energy exceeds a certain threshold~\cite{Barausse:2012da,Palenzuela:2013hsa}. Scalarization can only occur when $\beta_{0} < 0$, and in fact, Damour and Nordtvedt have shown that GR is an attractor in theory phase space when $\beta_{0} > 0$. Recently, Anderson et al~have shown that when $\beta_{0} < 0$, GR is a repeller in theory phase space, leading to a maximal violation of Solar System experiments, unless one fine-tunes the field's initial conditions at the time of Big Bang nucleosynthesis~\cite{Sampson:2014qqa,Anderson:2016aoi,Anderson:2017phb}. We will here ignore such problems, as done cavalierly in the literature.    

The degree of scalarization for isolated compact object $A$ is controlled by its scalar charge $\alpha_{A}$, which in general is a complicated function of $\alpha_{0}$, $\beta_{0}$ and the object's compactness. For WDs, one can approximate the scalar charge via
\be
\alpha_{A}^{\WD} = \alpha_0 \left[1 - 2 C_{A} + {\cal{O}}(C_{A}^{2}) \right]\,,  
\label{eq:alphaWD}
\ee
where $C_{A}$ is the star's compactness. The scalar charge is here independent of $\beta_{0}$ because these objects cannot become compact without collapsing into a black hole for the non-linear effects associated with $\beta_{0}$ to become relevant. For NSs, the scalar charge can only be calculated numerically by solving the equations of structure in scalar-tensor theories; we here use the tabulated expression of~\cite{david-private}.
 
Although Solar System observations cannot constrain $\beta_{0}$ since weakly gravitating objects have tiny scalar charges, the radio observation of binary pulsars can~\cite{Stairs:2003eg,Freire:2012mg}. These observations have accurately measured the rate at which the orbital period of certain binary pulsars decays, finding it to be in perfect agreement with GR. But when $\beta_{0}$ is sufficiently negative, then not only do NSs scalarize, but this activation of the scalar field also sources dipole radiation, which in turn forces the orbital period to decay more rapidly than in GR. This is precisely an example of the pre-Newtonian dissipation we have discussed previously. Given that such rapid decay is not observed, radio observations have placed constraints in the $(\alpha_{0},\beta_{0})$ plane. Due to degeneracies between these parameters in the timing model, only a certain combination of $(\alpha_{0},\beta_{0})$ can be bounded for any given observation. Constraints on the $(\alpha_{0},\beta_{0})$ plane from radio observations have already been presented in Fig.~\ref{fig:alpha_beta.png}.    

We use the same \LISA response function as described in Sec.~\ref{subsec:generic-modeling}. In principle, there is an additional scalar GW mode that ought to be included, but this is suppressed by the scalar charges, and we will thus ignore it here. The mapping between components of the model parameter vector $\vec{\lambda}$ is still given by Eq.~\eqref{eq:fdot-mapping} and second-order version of Eq.~\eqref{eq:fddot-mapping}, namely
\begin{align}
\label{eq:fddot-mapping2}
\fddot &= \fddot_{\GR} \left[1 
+ \frac{20}{11} \beta_{\ppE}  \left(\pi {\cal{M}} f\right)^{b_{\ppE}}  
+ \frac{9}{11} \beta_{\ppE}^{2} \left(\pi {\cal{M}} f\right)^{2 b_{\ppE}}
\right]\,,
\end{align}
with ${\cal{M}} \to {\cal{M}}_{*}$ and the particular choices 
\begin{align}
b_{\ppE} &= -2/3\,,
\label{eq:ppe0}
\\
\beta_{\ppE} &= \frac{5}{96} \eta^{2/5} \kappa^{-3/5} \left(1 + \alpha_{1} \alpha_{2}\right) \left(\alpha_{1} - \alpha_{2}\right)^{2}\,,
\label{eq:ppe}
\end{align}
where $\alpha_{1,2}$ are the scalar charges of the objects, while
\begin{align}
\kappa &= 1 + \frac{1}{6} \left(\alpha_{1} \frac{m_{2}}{m} + \alpha_{2} \frac{m_{1}}{m}\right)^{2} 
\nn \\
&+ \frac{1}{6} \left(\alpha_{1} - \alpha_{2}\right) \left(\alpha_{1} \frac{m_{1}}{m} + \alpha_{2} \frac{m_{2}}{m}\right) \left(\frac{m_{1}}{m} - \frac{m_{2}}{m} \right) 
\nn \\
&+ \frac{5}{48}  \frac{\alpha_{1} - \alpha_{2}}{1 + \alpha_{1} \alpha_{2}} \left(\beta_{2} \alpha_{1} \frac{m_{1}}{m} - \beta_{1} \alpha_{2} \frac{m_{2}}{m} \right) + d_{2} \left(\alpha_{1} - \alpha_{2}\right)^{2}\,.
\label{eq:kappa}
\end{align}
The quantity $\beta_{1,2}$ are the derivatives of the scalar charges, which we will ignore in this work for simplicity. The quantity $d_{2}$ is a numerical factor that has not yet been calculated, but that can be ignored because it is subdominant.  Notice that the quantity ${\cal{M}}_{*}$ in these equations is not quite the chirp mass, but rather it is the \emph{tensor chirp mass}, namely
\be
{\cal{M}}_{*} = \frac{\kappa^{3/5}}{\left(1 + \alpha_{1} \alpha_{2}\right)^{2/5}} {\cal{M}} \,,
\label{eq:mstar}
\ee
where we have set $G_{*} = 1$. 




\subsection{Constraints on Spontaneous Scalarization with $(\dot{f},\ddot{f})$ Measurements}

Consider a NS-WD system in the galaxy with GW frequency in the \LISA band.
While much rarer than the WD-WD systems, a NS-WD binary is a clearly identifiable source in the \LISA data stream due to its high mass, and will therefore be well localized ($d\Omega \ll 1$ sq. deg. after several years of observing) in the galaxy and yield a precise determination of the frequency evolution parameters~\cite{Cornish:2017vip}.
If electromagnetic follow-up observations constrain the masses and radii of the binary constituents, they would be used to predict the orbital evolution of the binary in scalar-tensor theories of gravity.
Those predictions, compared to the frequency parameters observed by \LISA, then serve as a test of the theory.

Exploring to what extent such a test yields novel insight into our understanding of gravity, we simulate a binary comprised of a 1.4 $\Msun$ NS and a 0.3 $\Msun$ WD with a GW frequency $\f = 10$ mHz near the galactic center, at a distance of 8 kpc (see also Table~\ref{tbl:wdwd}). 
Using the MCMC pipeline, we characterize how well the source parameters are measured. 
For electromagnetic follow up we would likely require the source be closer to Earth, and not towards the galactic center where EM observations will have to contend with extinction and source confusion. 
However, placing the sources near the galactic center is a conservative choice for the GW analysis: anything further away is difficult to observe electromagnetically, and anything closer results in better GW parameter estimation due to the increased \SNR. 

The marginalized posteriors from the MCMC analysis are approximated as multivariate Gaussians, as in Eq.~\eqref{eq:multivariate-gaussian} with parameters $\vec{\lambda} = (f_{0},\fdot,\fddot)$.
It is important to consider the full covariance matrix, rather than just the variance of the 1D marginalized posteriors for $\fdot$ and $\fddot$, because the two parameters are strongly correlated (see Fig.~\ref{fig:fdot_fddot.png}).
While the perils of blindly assuming Gaussian distributions for posteriors in GW astronomy are well documented, the sources we are analyzing have high \SNR ratios where this approximation is sufficient for proof of concept studies. 
Furthermore, we again confirm that the Gaussian approximation is adequate by testing that the distribution of p-values for the Markov chain samples evaluated using the Gaussian approximation to the posterior are uniformly distributed between 0 and 1, as we did in Sec.~\ref{subsec:example}. 

\begin{figure}
	\includegraphics[width=0.5\textwidth]{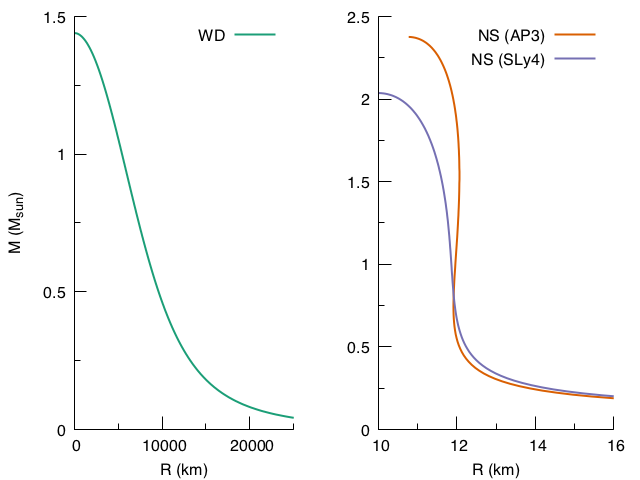}
	\caption{(Color online) Mass radius relationships used in this paper for the WD (left) and NS (right) models. The WD mass-radius relationship is a simple Newtonian approximation, which is adequate in the mass range we consider which is well below the maximum mass where relativistic effects become important.  The NS relationships come from numerical simulations assuming two equations of state: AP3~\cite{APR} and SLy4~\cite{SLy}. All results involving NSs in this work use the AP3 equation of state unless otherwise noted.}
	\label{fig:MR.png}
\end{figure}

As a stand in for the electromagnetic observations, we compute the WD and NS radii using mass-radius relations appropriate for the different compact objects.  For the WD we use a simple Newtonian relation, and for the NS we test two different equations of state, shown in Fig.~\ref{fig:MR.png}. Although currently the NS equation of state is not known accurately, by the 2030s when \LISA flies, additional LIGO/Virgo observations like the recent~\cite{TheLIGOScientific:2017qsa}, combined with observations of the X-ray pulse profile of bursting NSs by NICER~\cite{Ozel:2015ykl}, will have measured it much more accurately.  We then assume Gaussian errors for the masses of the objects and the radius, centered at the true value, testing assumed standard deviations of $10\%$, $1\%$, $0.1\%$, and $0.01\%$ of the true value. The NS radius is determined from the mass using the assumed equation of state.

\begin{figure*}
        \includegraphics[width=1.0\textwidth]{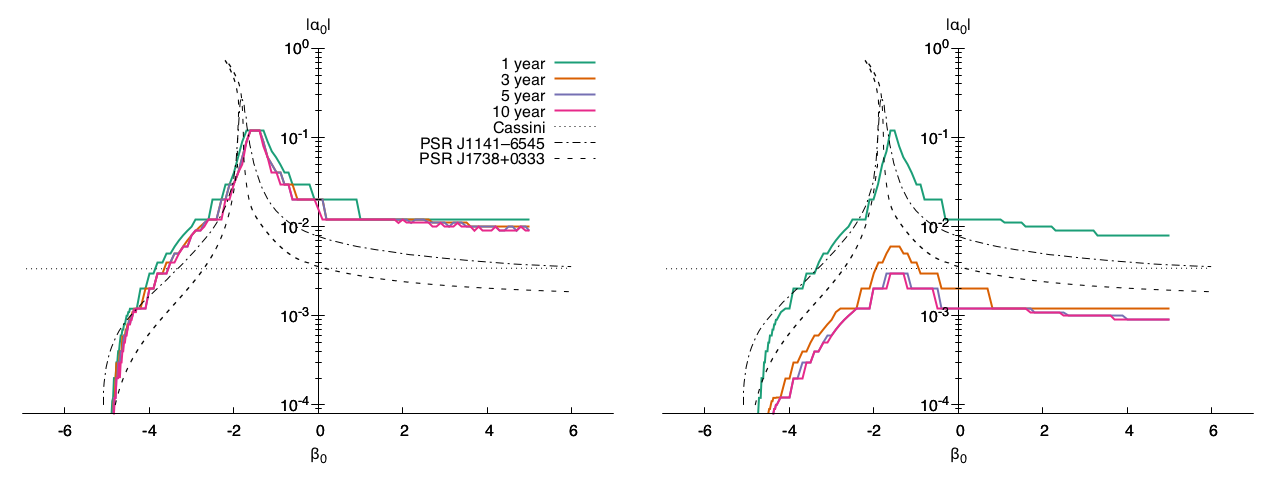}
        \caption{(Color online) Constraints on $\alpha_0$ vs. $\beta_0$ for canonical NS-WD system at different observing times. The left (right) panel assumes 1\% (0.01\%) errors on mass and radius measurements. The mass/radius uncertainties dominate the bounds on scalar-tensor parameters over the GW measurement until they are reduced to the extremely optimistic $0.01\%$ level. At that point we see the bounds improve with observing time, leading to better than current constraints.}
        \label{fig:alpha_beta_bound.png}
\end{figure*}

In lieu of a closed-form expression for mapping the EM and GW parameters to scalar-tensor theory parameters, we test whether a set of discrete values for $\alpha_0$ and $\beta_0$ are jointly consistent with $p(\fdot, \fddot \vert d)$ approximated from the MCMC, marginalizing over the inferred $\f$ measurement from the GW data and the assumed mass and radius uncertainties from EM observations.
We begin by randomly drawing $10^6$ combinations of $\f$, masses, and radii for the NS and WD consistent with $p(\f\vert d)$ and the assumed EM measurement capabilities, respectively.
For each set of masses and radii, we compute the scalar charge of the WD using Eq.~\eqref{eq:alphaWD}. 
The NS scalar charge is obtained from a look up table of numerical simulations on a discretized grid of mass, $\alpha_0$, and $\beta_0$~\cite{david-private}. Then, using Eqs.~\eqref{eq:ppe0}-\eqref{eq:mstar}, and substituting into Eqs.~\eqref{eq:fdot-mapping} and \eqref{eq:fddot-mapping2}, we calculate the predicted $\fdot^*$ and $\fddot^*$ for that combination of parameters.  
If the $p$-value of the marginalized posterior $p(\fdot, \fddot \vert d)$ evaluated at the predicted $\fdot^*$ and $\fddot^*$ is greater than $0.1$ the scalar tensor parameters are considered consistent with the GW and EM measurements, from which upper limits are placed on $\alpha_0$ as a function of $\beta_0$.

The bound that can be set using joint EM+GW observations is heavily dependent on how well the masses and radii of the binary components can be independently measured by EM observations.
Upper bounds at 90\% confidence for a ten year observation of our canonical NS-WD binary system, assuming mass and radius errors ranging from $10\%$ to (an admittedly optimistic) $0.01\%$, are compared to existing bounds in Fig.~\ref{fig:alpha_beta.png}.  
For \LISA observations to improve on existing bounds of scalar-tensor theories, long observing times and independent precision measurement of the mass and radii of the binary are necessary.

To see if this conclusion is generically true, or if it depends on properties of the source, we repeat the analysis for different observing times, orbital periods of the binary, WD mass, mass ratio, and NS equation of state. 
Below, we present the $1\%$ and $0.01\%$ accuracy cases for EM-derived mass and radius constraints as representative of plausible and best case outcomes, respectively.

Figure~\ref{fig:alpha_beta_bound.png} shows how the \LISA constraints improve over the mission lifetime.
For systems with mass and radius constraints in the $1\%$ range (left panel), the uncertainty in the properties of the binary constituents dominates the constraint on $\alpha_0$, and we see minimal improvements even as the orbital evolution parameters become increasingly well measured with longer observing times.
Adopting the optimistic mass and radius constraints (right panel), the bounds become competitive, and ultimately exceed, existing bounds from \emph{Cassini} and observations of binary pulsars.
Because we are assuming joint electromagnetic observations, these systems could be monitored well beyond the \LISA mission lifetime, perhaps with follow-on GW observatories, to further improve the bounds.

We also investigate how the orbital period of the binary affects the bounds that can be inferred, which presents an interesting trade space to explore. 
Dipole radiation plays a more prominent role in the orbital evolution at lower orbital velocities, i.e. low GW frequency.
However, \LISA will have less sensitivity at lower frequencies due to the presumed levels of confusion noise from unresolved binaries in the galaxy. 
Furthermore, measuring frequency derivatives is more difficult at low frequency because the overall change to the GW phase during the observing time is smaller.

\begin{figure*}
	\includegraphics[width=1.0\textwidth]{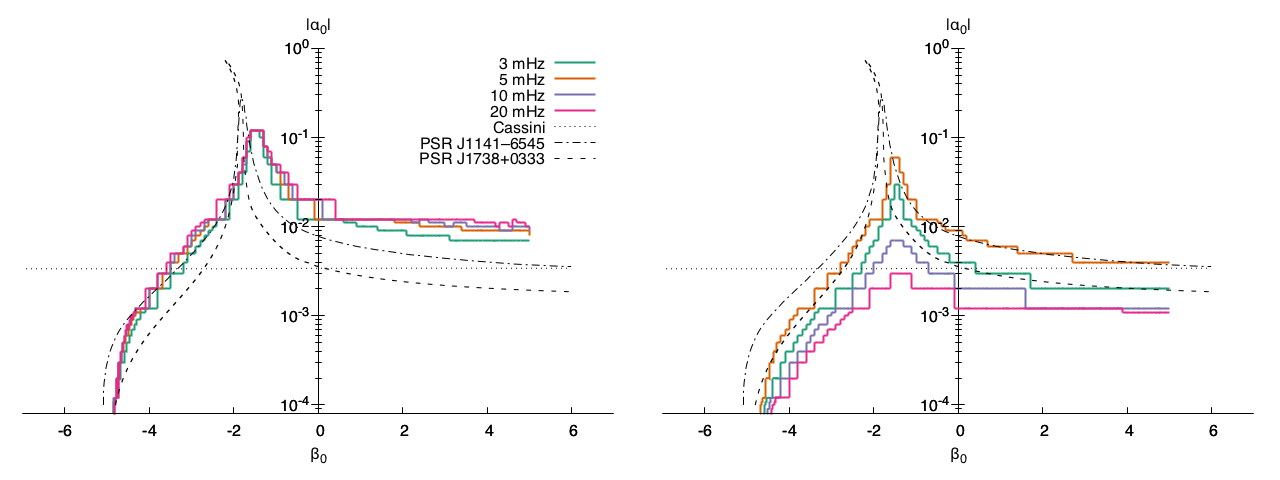}
	\caption{(Color online) Same as Fig.~\ref{fig:alpha_beta_bound.png} but for systems at different GW frequencies. The mass/radius uncertainties dominate the bounds on scalar-tensor parameters over the GW measurement until they are reduced to the extremely optimistic $0.01\%$ level. With high precision mass radius measurements, we see that the high frequency binaries provide the best constraints, although that is not monotonically true, as the bound at $3\ {\rm mHz}$ exceeds that at $5\ {\rm mHz}$ because the dipole radiation effects are more prominent at low velocities.}
	\label{fig:alpha_beta_bound_frequency.png}
\end{figure*}

We test which of these effects wins out by comparing bounds derived for the $1.4+0.3\Msun$ NS-WD system at different GW frequencies ($\f=3,5,10$ and $20\ {\rm mHz}$) in Fig.~\ref{fig:alpha_beta_bound_frequency.png}. Observe how we again find, for $1\%$ uncertainties, that the mass/radius measurement dominates the bound on scalar-tensor parameters.
If EM constraints could reach the $.01\%$ level, we see separation of the bounds depending on the frequency of the binary, with the higher frequency systems providing tighter bounds owing to the better measurement of $\fdot$ and $\fddot$.  
Interestingly, the bound does not monotonically improve with frequency, as the $3\ {\rm mHz}$ source better constrains $\alpha_0$ than the $5\ {\rm mHz}$ system.

\begin{figure*}
	\includegraphics[width=1.0\textwidth]{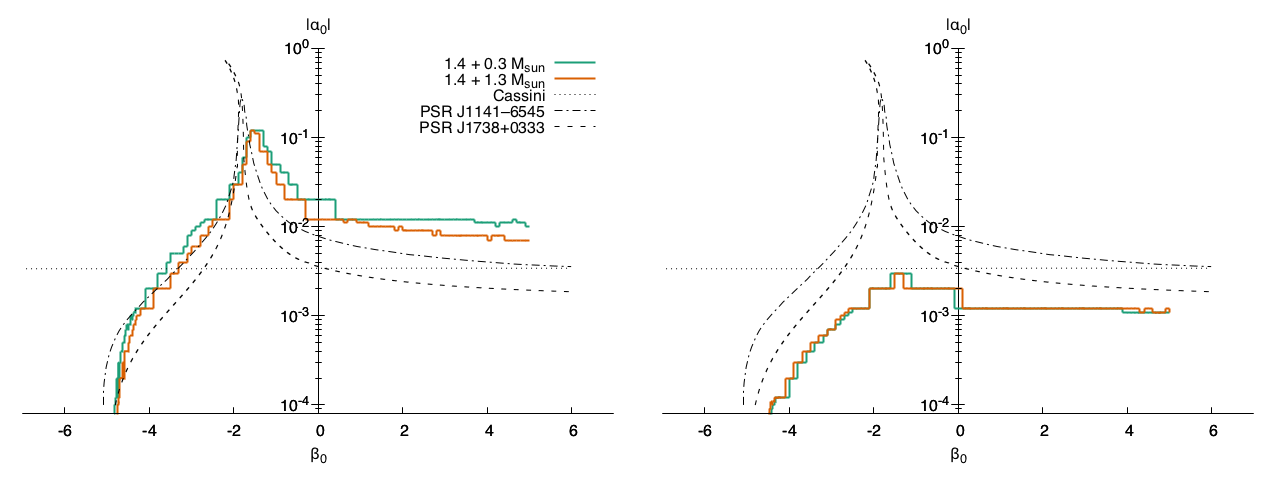}
	\caption{(Color online) Same as Fig.~\ref{fig:alpha_beta_bound.png} for canonical NS-WD systems with varying component masses. We see no discernible difference in the bounds for different mass binaries.}
	\label{fig:alpha_beta_bound_mass.png}
\end{figure*}

The final result presented here explores how the mass of the WD companion affects the bounds on the scalar-tensor parameters. The reason for this study is that a higher mass WD would be more compact, and therefore would have larger scalar charge, which, in turn, would be a stronger source of dipole radiation.
Alas, we find little difference in the bounds derived from the GW measurement comparing $1.4+0.3\Msun$ and $1.4+1.3\Msun$ NS-WD binaries at $\f=20\ {\rm mHz}$, as we can see in Fig.~\ref{fig:alpha_beta_bound_mass.png}. We also investigated the effect of the NS composition, comparing bounds using the AP3~\cite{APR} and SLy4~\cite{SLy} equations of state, and found no measurable difference between the derived bounds.

\section{Caveats}
\label{sec:caveats}

In this section we present several caveats on our analysis that could be investigated further in the future. We begin by discussing the dependence of our results on the galactic binary population. We end with a short discussion of the effects of tidal and/or rotational deformations. 

\subsection{Dependence on the Population}
For the WD binary sources, we are using a small sample of the shortest period binaries in a simulated galaxy.
These have shorter periods than any known binaries at the time of this study, and so there existence is a prediction of population simulations.  
From those simulations, there will be a large amount of variance in this high mass, short period sub-population, and the sample of such binaries in the observed galaxy may not be as kind.
That being said, current observations of binaries are limited to nearby, bright sources, whereas sources with these masses and GW frequencies will be observable throughout the galaxy so, if they exist, they will be excellent \LISA sources.
A similar argument exists for NS-WD binaries in the \LISA band.
Further population synthesis studies, and electromagnetic surveys searching for short period binaries, may strengthen or weaken the case for these extraordinary systems.
It may also be the case that we will just have to wait to see what \LISA observations have in store.

\subsection{Tidal and/or Rotational Deformations}
In this work we have intentionally isolated a particular modification to the binary evolution (dipole radiation) to determine how well it can be constrained using \LISA observations.
The effect of dipole radiation will be competing with other, similarly difficult to measure, processes that will modify the phase evolution of the binary from the point-particle-in-GR case.
Mass transfer is an obvious culprit, though such binaries are expected to be a subset of the total observed population, particularly at the high end of the $\Mc$ distribution, where our simulated sources are found.

Tidal interactions between the stars, however, may prove to play an important role generically, especially for the high frequency binaries we consider here.
Fortunately for us, tidal contributions enter at much higher post-Newtonian order. More specifically, the tidal deformability enters at $5$ post-Newtonian order, and thus, its effects would scale as $(\pi {\cal{M}} f_{0})^{21/3}$ in Eq.~\eqref{eq:fdotGR} and $(\pi {\cal{M}} f_{0})^{29/3}$ in Eq.~\eqref{eq:fdotGR2}. These modifications are multiplied by a tidal deformability parameter that for WDs is very large, as it scales as $C^{-5}$ where $C$ is the gravitational compactness, and that is enhanced by non-zero orbital eccentricity~\cite{Willems:2008,Willems:2009xk,Piro:2011qe,Valsecchi:2011mv}. However, an order of magnitude estimate suggests that the 5PN suppression is at least 5 orders of magnitude stronger than the tidal deformability parameter enhancement, leading to an overall effect that should be very small. Ultimately, an holistic approach to modeling the phase evolution of galactic binaries in \LISA data including these different effects will be needed to avoid biasing inferences about any one potential source of orbital evolution.

\section{Conclusions}
\label{sec:conclusions}

We studied the potential of using galactic binary GW observations with LISA to carry out tests of GR. In spite of the current lore, there are a large class of modified theories of gravity that predict large modifications to the temporal or spectral evolution of such GWs. We thus developed a model-independent framework to carry out null tests of GR with such GWs, which resemble those carried out with radio observations of binary pulsars.  We performed several MCMC studies to determine the strength of such null tests, and found that long observing times (of order $5$--$10$  years) would be most probably required.

We then investigated model-dependent test of GR with future LISA observations of galactic binaries. In particular, we focused on tests of scalar tensor theories that predict dipole emission in compact binaries. We performed several MCMC studies to determine the strength of such model-dependent tests, and found that shorter observing times (of ${\cal{O}}(3$--$5)$ years) would suffice to place better than current constraints. The caveats here though are that such model-dependent tests require an observation of a mixed NS-WD galactic binary and knowledge of the radius of the NS, which could be obtained with coincident electromagnetic observations if the binary is eclipsing.            

We also investigated whether modifications to GR in the signal could affect the astrophysical inferences one would obtain from LISA observations of galactic binaries analyzed with GR templates. We found that indeed fundamental bias can be present, inducing a systematic error in the extracted GR parameters of ${\cal{O}}(10\%)$, even before a deviation is measurable. Such stealth bias eventually disappears as the observing time increases and the deviations begin to induce a failure of the null tests we developed. This implies that accurate chirp mass inferences with GR templates may require marginalizing over these non-GR effects to within existing bounds.      

We hope the analysis presented here catalyzes an explosion of future work. One issue that is particularly important is the impact of astrophysical systematics on the model-independent and model-dependent tests considered here. Perhaps the most important astrophysical effect are tidal deformations on WDs, which would induce a correction on the waveform template at high post-Newtonian order. These corrections should be taken into account to determine whether they can spoil the tests discussed here.  
 
Another possible avenue for future work is the study of follow-on missions beyond LISA. Even after LISA flies and is de-commissioned, electromagnetic observing campaigns of the Milky Way will continue. Long enough observations could then allow for measurements of the radii of eclipsing NS-WD galactic binaries, even of these radii measurements are not available when LISA flies. Once this is done, one could then go back to the LISA data to carry out the model-dependent tests described here. Moreover, follow-on space-based GW missions, like the Big Bang observer, will observe galactic binaries later in their orbital evolution, and at much higher S/N, leading to even stronger constraints. It would be interesting to investigate the level to which such tests could be carried out.  

The analysis described above, however, requires knowledge of the equation of state. By the time LISA flies in the 2030s, one expects the equation of state to have been constrained significantly with future observations by LIGO/Virgo and NICER. However, still some level of uncertainty will remain, which would have to be marginalized over to carry out the model-dependent tests described here. It would be interesting to carry out such a marginalization analysis to determine how the strength of the model-dependent tests would be affected.    

\acknowledgments We would like to thank Neil Cornish for useful comments and discussion. We are also most appreciative to David Anderson, who provided thorough tables for the scalar charges of NSs as a function of compactness and the coupling constants $(\alpha_{0},\beta_{0})$. 
N. Y. acknowledges support from NASA grants 4W5883, NNX16AB98G and 80NSSC17M0041. 
T. L. was supported by NASA grant NNH15ZDA001N and the NASA \LISA Study Office.

\bibliography{master}

\end{document}